# Response to "Comment on 'Widely tunable compact terahertz gas lasers'"

Paul Chevalier[1], Arman Amirzhan[1], Fan Wang[2], Marco Piccardo[1], Steven G. Johnson[3,4], Federico Capasso[1] and Henry O. Everitt[5,6]

[1]Harvard John A. Paulson School of Engineering and Applied Sciences, Harvard University, Cambridge, MA 02138, USA.
[2]Department of Mechanical Engineering, Massachusetts Institute of Technology, Cambridge, MA 02139, USA.
[3]Department of Mathematics, Massachusetts Institute of Technology, Cambridge, MA 02139, USA.
[4]Department of Physics, Massachusetts Institute of Technology, Cambridge, MA 02139, USA.
[5]U.S. Army Combat Capabilities Development Command Aviation and Missile Center, Redstone Arsenal, AL 35898, USA.
[6]Department of Physics, Duke University, Durham, NC 27708, USA

**Abstract:** We recently demonstrated a widely tunable THz molecular laser and reported mathematical formulas and a table for comparing how various molecules would perform as such lasers (Chevalier *et al.*, Science, 15 November 2019, p. 856-860). Here we correct the value of a single parameter used to calculate the table (see Erratum for Chevalier *et al.*), thereby eliminating the concerns raised by Lampin and Barbieri (Lampin *et al.*, arXiv:2004.04422). We also show that our simplified model for the output THz power is a better approximation than the alternative one proposed in the technical comment.

## I. Correction of a single parameter eliminates the concerns raised by Lampin and Barbieri

In our paper [1], we proposed that quantum cascade laser (QCL)-pumped molecular lasers (QPMLs) can lase on virtually any rotational transition in most molecular gases, including a few ammonia inversion transitions as initially proposed in [2]. The continuous tunability of QCLs makes this possible for virtually any gaseous molecule with a permanent dipole moment. A comprehensive ab-initio model [1,3] was used for predicting and comparing with the experimental measurements, but we also employed a greatly simplified model, valid only at low pressures, for a specific purpose: comparing the THz lasing performance and tunability of several different molecules. That simple model formula

$$P_{THz} = \frac{T}{4} \frac{\nu_{THz}}{\nu_{IR}} \frac{\alpha_{IR}}{\alpha_{cell}} \left(P_{QCL} - P_{th}\right) = \eta(P_{QCL} - P_{th}) \qquad (1)$$

shows how the output power $P_{THz}$ from a given rotational transition of frequency $\nu_{THz}$ of a candidate molecule, pumped by a tunable QCL with frequency $\nu_{IR}$, depends on the infrared absorption of the pumped ro-vibrational transition $\alpha_{IR}$, the total gas cell losses (ohmic and mirror losses) $\alpha_{cell}$, the output coupler transmission $T$, the pump power $P_{QCL}$, and the threshold power $P_{th}$, for which an expression was also presented in the original paper [1]. Note that the Manley-Rowe limit ($\eta = \frac{\nu_{THz}}{\nu_{IR}}$) is not violated as long as $\frac{T}{4} \frac{\alpha_{IR}}{\alpha_{cell}} < 1$, which is almost always true in our compact QPML for pressures below 20 mTorr.

To see this, we address a concern expressed by the authors of [4] about the value of $\alpha_{cell} = 0.06\ m^{-1}$ given in [1] for the ideal cavity in Table 1. We computed this value by estimating the ohmic losses as $\alpha_{Ohmic} = 0.01\ m^{-1}$ and a transmission coefficient $T = 0.016$ for a 15 cm long cavity. The value $T = 0.016$ was computed from the overlap integral between the $TE_{01}$ mode and the 1 mm diameter pinhole coupler in the 5 mm diameter cavity. Therefore, we did in fact include the outcoupling pinhole losses when computing $\alpha_{cell}$, as opposed to the claim of [4]. However, in using Eq. 1 to produce the values of Table 1 in [1] we mistakenly used $T = 0.04$, computed as the square of the ratio of the pinhole coupler radius and the cavity radius unweighted by the overlap with the $TE_{01}$ mode.

Here, and in the Erratum for [1], we include a corrected table where the value of the output coupler transmission ($T = 0.016$) is now the same as the one used to compute $\alpha_{cell} = 0.06$ m$^{-1}$. This has reduced $\eta$ and $P_{THz}$ by a factor of 2.5 for every molecule without changing any of our conclusions. More importantly, it eliminates the concern about violating the Manley-Rowe limit, which occurs only if $\alpha_{IR} > 15$ m$^{-1}$ in the ideal cavity and $\alpha_{IR} > 75$ m$^{-1}$ in the lossy cavity we used ($\alpha_{cell} = 0.3$ m$^{-1}$). These values of $\alpha_{IR}$ are larger than the one considered in Table 1 of the paper for N$_2$O ($\alpha_{IR} = 12.7$ m$^{-1}$).

To avoid this concern entirely, we may go beyond the assumption that the infrared absorption is small (i.e. $\alpha_{IR} L \ll 1$), where L is the cavity length. This is the limiting case of the more general expression

$$\alpha_{IR} = \frac{1 - exp(-\alpha_0 L)}{L} \qquad (2)$$

where $\alpha_0$ is the absorption coefficient for the infrared ro-vibrational transition pumped by the QCL. The approximation used in the simple model reduces Eq. 2 to $\alpha_{IR} \approx \alpha_0$, but this assumption can be removed by using the exact value from Eq. 2 in Eq. 1 without loss of generality. As pointed out by [4], the largest value $\alpha_{IR}$ may achieve (when $\alpha_0 L \gg 1$) is $1/L = 6.7$ m$^{-1}$ for our cavity, far from violating the Manley Rowe limit.

The updated table, using the corrected value of $T$ and Eq. 2 for $\alpha_{IR}$, is below. When sorted by increasing $P_{th}$ as before in [1], there is little change in the original ordering. Instead, the largest change is in the line $J_L$ and frequency $\nu_{THz}$ producing the most power, both of which have generally increased because of the use of Eq. 2. Compared to [1], and since Eq. 2 is used, the table now gives the value of $\alpha_0$ instead of $\alpha_{IR}$.

| Molecule | $J_L$ (peak) | $\nu_{THz}$ (THz) | $P_{THz}$ (mW) | $P_{th}$ (mW) | $\eta$ (mW/W) | $\nu_{IR}$ $(cm^{-1})$ | $\alpha_0$ $(m^{-1})$ |
|---|---|---|---|---|---|---|---|
| CH$_3$F | 17 | 0.907 | 1.40 | 0.063 | 5.60 | 1075.379 | 3.96 |
| NH$_3$ | 3 | 1.073 | 3.3 | 0.109 | 13.2 | 967.346 | 10.9 |
| HCN | 11 | 1.064 | 0.29 | 0.178 | 1.14 | 1447.962 | 0.74 |
| H$_2$CO | 12 | 0.970 | 0.28 | 0.229 | 1.12 | 1776.936 | 0.99 |
| OCS | 43 | 0.535 | 0.77 | 0.251 | 3.10 | 2077.629 | 11.1 |
| CH$_3{}^{35}$Cl | 21 | 0.585 | 0.015 | 2.43 | 0.06 | 1459.576 | 0.068 |
| CH$_3$OH | 15 | 2.523 | 0.090 | 4.90 | 0.37 | 1031.477 | 0.068 |
| N$_2$O | 27 | 0.703 | 0.83 | 8.08 | 3.43 | 2244.404 | 8.92 |
| CO | 10 | 1.268 | 0.88 | 41.2 | 4.19 | 2183.224 | 4.45 |

## II. The simple model of our paper is a better approximation than the alternative proposed in [4]

The restriction of this simple model to very low pressures (typically ≤ 20 mTorr) derives from the assumption that the dominant molecular relaxation mechanism in compact QPML cavities is by ballistic wall collisions rather than by dipole-dipole collisions. For higher pressures, a comprehensive model has been developed, but it is much more computationally intensive and requires the knowledge of additional molecular collision cross sections (dipole-dipole, gas kinetic, vibrational state change) that may not be known [1,3]. Thus, by using the simple model in the low-pressure regime where these parameters are not needed, we may ascertain a qualitative comparison of how a given molecular gas may perform in a QPML. Moreover, the simple model allows one to estimate the broad tunability that may be achieved from a given molecule by estimating an upper bound on how $P_{THz}$ varies from line-to-line, a very useful capability given that virtually any rotational transition that can absorb terahertz radiation may operate as a laser on that same frequency. Most importantly, our simple model reveals that the threshold pump power $P_{th}$ for many lines in most molecules in a compact QPML is well below what a typical QCL can emit, confirming the universality of this concept.

By contrast, the approximation proposed in [4], $\alpha_{IR} \approx 1/L$, cannot provide these insights because it removes the molecule-specific, line-specific dependence of $\eta$ and $P_{THz}$ on $\alpha_0$, so it is unable to provide the performance comparisons for which the simple model was designed. In that limit, the only variation in $P_{THz}$ from line to line and molecule to molecule comes from the terms $\nu_{THz}$, $\nu_{IR}$, and $P_{th}$. Even if such an approximation is acceptable for operating a QPML over a narrow range of frequencies and pressures, as was done in previous work [2,5], it is not valid for estimating how the widely tunable QPML performance varies from molecule-to molecule and from line-to-line. Moreover, the approximation proposed in [4] requires that the pressure and/or cavity length be varied to achieve the $\alpha_{IR} L > 1$ condition. For the 20 mTorr case presented in the revised table 1 above, this would require the "compact" laser cavity length to vary from $1/11.1$ m$^{-1}$ = 9 cm for OCS to $1/0.068$ m$^{-1}$ = 15 m for CH$_3$Cl and CH$_3$OH! Such length variations are impractical and often not compact. And as noted above, this approximation always overestimates the performance of a QPML because $1/L$ is the largest value of $\alpha_{IR}$ any transition or molecule can have.

The approach that we proposed in our work instead focuses on a practical implementation of the QPML concept where the laser cavity is kept at the same short length and tuning is achieved by pumping different infrared transitions then adjusting the cavity length (by much less than a millimeter) to match a cavity mode to the lasing transition. In the experiment we have used the same short (15 cm long) laser cavity, filled with the same gas, pumped by the same QCL, at varying pressure between 20 and 50 mTorr to generate emission from 250 GHz to 950 GHz. The infrared absorption of the gain medium depends on the infrared frequency at which the QPML is pumped, which in turn determines the output terahertz frequency. Therefore, it is necessary to take the IR absorption $\alpha_0$ into account in order to understand how the performance of such a laser changes with output frequency and cavity pressure.

## III. Conclusion

The core of our paper [1] described an experimental proof-of-concept realization of a broadly tunable QPML using nitrous oxide, which lased on every line between 250 GHz and 950 GHz, confirming the key predictions of the simple model. In addition, the performance of this laser was compared to predictions by the comprehensive model at various pump powers and pressures, including pressures too high for the simple model to be used. Extrapolating this simple model beyond its realm of validity, such as the authors of [4] attempted to do by using it to analyze our measurements on $N_2O$ lasing at 40 mTorr, is inappropriate and may lead to incorrect predictions and even unphysical results.

We reiterate that the authors of the technical comment [4] did not challenge the core conclusions of our paper [1] that any molecular gas with a permanent dipole moment can be made to lase on virtually any transition when pumped by a QCL tuned across one of its vibrational bands. They did not challenge our experimental findings or methods either. The simple model derived in our paper [1] is correct when used in the very low pressure regime and is the most efficacious way to understand the molecule- and line-dependent behavior of a compact QPML. While it is not appropriate to use this simple model outside the very low pressure regime, we point out more significant limitations in the alternative approximation proposed in [4] and acknowledge that the comprehensive model presented in [1,3] supersedes both the simple model (Eq. 1) and that of the technical comment [4].

**References**
[1] P. Chevalier *et al.*, *Science*, **366**, 856-860 (2019).
[2] A. Pagies, G. Ducournau, J.-F. Lampin, *APL Photonics* **1**, 031302 (2016).
[3] F. Wang *et al.*, *Proc. Natl. Acad. Sci. U.S.A.* **115**, 6614–6619 (2018).
[4] J.-F. Lampin, S. Barbieri, Comment on "Widely tunable compact terahertz gas lasers", arXiv:2004.04422 (2020).
[5] J.-F. Lampin *et al.*, Opt. Express, **28**, 2091 (2020).